\documentclass[twocolumn]{revtex4}
\usepackage{graphicx}
\usepackage{amsmath}
\usepackage{amssymb}
\usepackage{epsfig}
\newcommand{\be}{\begin{equation}}
\newcommand{\ee}{\end{equation}}
\newcommand{\ba}{\begin{eqnarray}}
\newcommand{\ea}{\end{eqnarray}}

\newcommand{\gev    }{\ensuremath{\mathrm{GeV}}}
\newcommand{\gevsq  }{\ensuremath{\mathrm{GeV^2}}}

\begin{document}
\title{Comparison of a new $\Delta$ resonance production model 
with electron and neutrino data}

\author{
E.\ A.\ Paschos$^1$, M.\ Sakuda$^2$, 
I.\ Schienbein$^3$, \underline{J.\ Y.\ Yu}$^1$
}
\vspace{1.5cm}
\affiliation{$^1$ Theoretische Physik III, University of Dortmund, 
44221 Dortmund, Germany
\\
$^2$ Department of Physics, Okayama University, Okayama, 700-8530
Japan
\\
$^3$ Deutsches Elektronen Synchrotron (DESY), Notkestrasse 85, 
22603 Hamburg, Germany
\\
Presented by J. Y.\ Yu at NUINT04, March, 2004, Gransasso, Italy
}

\begin{abstract} 
We consider resonance production by neutrinos focusing on the dominant 
resonance $P_{33}$ at low energies with a detailed discussion 
of its form factors. The results are presented for free nucleon targets.
The $\Delta$ resonance is described by two 
form factors $C_3^V$ and $C_5^A$ and its differential cross sections 
are compared with experimental data.
Further, we apply this approach to the electroproduction case 
and calculate its differential cross sections which 
are compared with electroproduction experimental data.     
Our approach to the analysis of $\Delta$ resonance 
is particularly simple and self-contained such that it could be 
helpful for the physical interpretation.
\end{abstract} 

\maketitle 
 
\section{Introduction} 
The excitation of the resonances by electrons and neutrinos 
has been studied extensively for a long time.
The earlier articles \cite{ref1,ref2,ref3,ref4,ref5} tried to determine 
the $p\Delta$ transition form factors in terms of basic principles, 
like conserved vector current (CVC), 
partially conserved axial-vector current (PCAC), dispersion relations, etc.
These and subsequent papers introduced dipole form factors and 
in various cases other functional forms with additional kinematic 
factors in order to reproduce the data. 
As a result, the cross sections (differential and integrated) were presented 
in terms of several parameters \cite{ref6,ref7,ref8,ref9}.
The relatively large number of parameters and the limited statistics of 
the experiments provided qualitative comparisons but an accurate 
determination of the terms is still missing.
A new generation of experiments is now under constructions aiming to measure 
properties of the neutrino oscillations and they will provide 
the opportunity precise tests of the standard model.

For these reasons it is important to improve 
the calculation of the excitation of resonances with isospin $I=3/2$ 
and $I=1/2$ looking into various terms that enter the calculations 
and trying to determine them, as accurately as possible.
This has been done recently by three of us \cite{ref10}.
In this contribution we will focus on the excitation of the 
$\Delta$ resonance which is dominant at the low energies
considered here \cite{ref10} and which is relevant for the study of
single pion production in current and future long baseline (LBL)
experiments, like K2k, MiniBoone, MINOS, CERN-GS, JHF, etc.
We give explicit formulas and discuss form factors 
for $P_{33} (1232)$ 
where
the formulas and form factors for other resonances 
$P_{11} (1440),\ S_{11} (1535)$ and $D_{13} (1520)$ can be found 
in \cite{ref10}.

In order to enlarge the event rates
neutrino experiments use medium-heavy and heavy nuclei which brings in
additional corrections such as the Pauli exclusion principle, 
Fermi motion and absorption and charge exchange of the produced pions 
in nuclei. 
However, in this contribution we restrict ourselves to free nucleon
targets and refer to refs.\ \cite{ref11} for a discussion of
nuclear effects included in our approach.

The paper is organized as follows.
In Sec.\ \ref{del} we give the general formalism for the production
of the $\Delta$ resonance following \cite{ref10}
emphasizing the minimal input, which is necessary.
In Sec.\ \ref{neu} we present numerical results of our
formalism for neutrino production of the $\Delta^{++}$.
In addition to the results presented
in \cite{ref10} 
we compare with experimental results for the
ratio of single pion resonance production (RES) and 
quasi elastic scattering (QE) which is important for understanding the
problematic low $Q^2$ region.
Furthermore,
we include several additional
consistency checks of the theory by comparing with 
more data for the $Q^2$ spectrum of single pion neutrino production.
In Sec.\ \ref{ele} we turn to a first comparison of our simple approach
with (fully differential) electroproduction data from JLAB
and finally draw our conclusions in Sec.\ \ref{summary}.

\section{Neutrino production of $\Delta$ resonance: Formalism} \label{del} 

The double differential cross section for $\Delta$ resonance production is 
given by  
\begin{eqnarray}
\frac{{\rm d} \sigma}{{\rm d} Q^2 {\rm d} W^2} =
\frac{G_F^2}{16\pi M^2}\sum_{i=1}^{3} [K_i W_i]\label{eq:douff}
\end{eqnarray}
with $G_F$ the Fermi constant and M the nucleon mass.
The kinematic factors $K_i(Q^2,E_\nu,W)$ 
and the structure functions $W_i(Q^2,W)$ which are expressed in terms of 
helicity amplitudes are given in Ref. \cite{ref5}.  
The helicity amplitudes depend on the Breit-Wigner factor 
\begin{eqnarray}
f(W) = \frac{\sqrt{\frac{\Gamma_\Delta(W)}{2\pi}}}{(W-M_\Delta)
-\frac{1}{2} i\Gamma_\Delta(W)} \label{eq:2}
\end{eqnarray} 
with $\Gamma(W) =\Gamma_0\ q_\pi(W)/q_\pi(W_R)$ and 
$\Gamma_0 = 120\ {\rm GeV}$
and form factors $C_i^V,C_i^A, i = 1,...,5$ 
which are discussed in the following.

For the matrix element of the vector current we use
the general form
\begin{eqnarray}
\langle \Delta^{++}|V_{\mu}|p\rangle &=&
\left[ \bar{\psi}_{\mu} A_{\lambda}q^{\lambda}-
\bar{\psi}_{\lambda}q^{\lambda} A_{\mu}+C_6^V(q^2)
\bar{\psi}_{\mu}\right] \times \nonumber \\
&& \gamma_5 \,u(p)\,f(W)
\end{eqnarray}
with $A_{\lambda}=\frac{C_3^V}{M}\gamma_{\lambda}
+\frac{C_4^V}{M^2} p'_{\lambda}+\frac{C_5^V}{M^2}p_{\lambda}$,
$C_i^V$, $i=3, \ldots, 6$ the vector form factors, 
$\psi_{\mu}$ the 
Rarita--Schwinger wave function of the $\Delta$--resonance,
and $f(W)$  the $s$--wave Breit--Wigner resonance,
given explicitly in eq.\ (\ref{eq:2}).
The conservation of the vector current (CVC) gives
$C_6^V(q^2)=0$ and the other form factors are determined
from electroproduction experiments, where the magnetic
form factor dominates. This dominance leads to the 
conditions
\begin{equation}
C_4^V(q^2) = -\frac{M}{W}C_3^V(q^2)\quad\quad
              {\rm{and}} \quad\quad C_5^V(q^2) 
           = 0.
\label{eq:c4v}
\end{equation}
With these conditions the electroproduction data depend
only on one vector form factor $C_3^V(q^2)$.
Precise electroproduction data determined the form
factor, which can be parameterized in
various forms.

Early articles describe the static theory \cite{ref13} and the quark model
\cite{ref14} predicting the form factor for 
the $\gamma N \Delta$ vertex to be proportional to the isovector part of 
the nucleon form factors. Subsequent data \cite{ref15} 
showed that the form factor for $\Delta$-electroproduction 
falls faster with increasing $Q^2$ than the nucleon form factor 
which motivated some authors to introduce 
other parameterizations including exponentials \cite{ref16} 
and modified dipoles \cite{ref17}.
The functional form
\begin{equation}
C_3^V(Q^2) = \frac{C_3^V(0)}{\left[1+\frac{Q^2}{M_V^2}\right]^2}
              \frac{1}{(1+\frac{Q^2}{4 M_V^2})} 
\label{eq:ff}
\end{equation} 
gives an accurate representation. 
In our approach \cite{ref10} we adopt this vector form factor 
and use CVC to determine its contribution to the neutrino induced reactions.
We can see in fig.\ \ref{tiator} that 
our modified form factor in eq. (\ref{eq:ff})
agrees well with resent experimental results for the 
magnetic $N-\Delta$ transition form factor $G_M^\star$ from
Mainz, Bates, Bonn and JLab \cite{tit}.
%We also agree with the Sato-Lee form factor which has been shown to be
%consistent with new JLab measurements of $G_M^\star$ \cite{sl}.
Details of the vector and axial contributions 
are presented in section \ref{neu}, 
where we shall estimate the contribution 
of $C_3^V$ from the electroproduction data.
\begin{figure}[htb]
\centering
\vspace*{-0.22cm}
\includegraphics[angle=0,width=8.5cm]{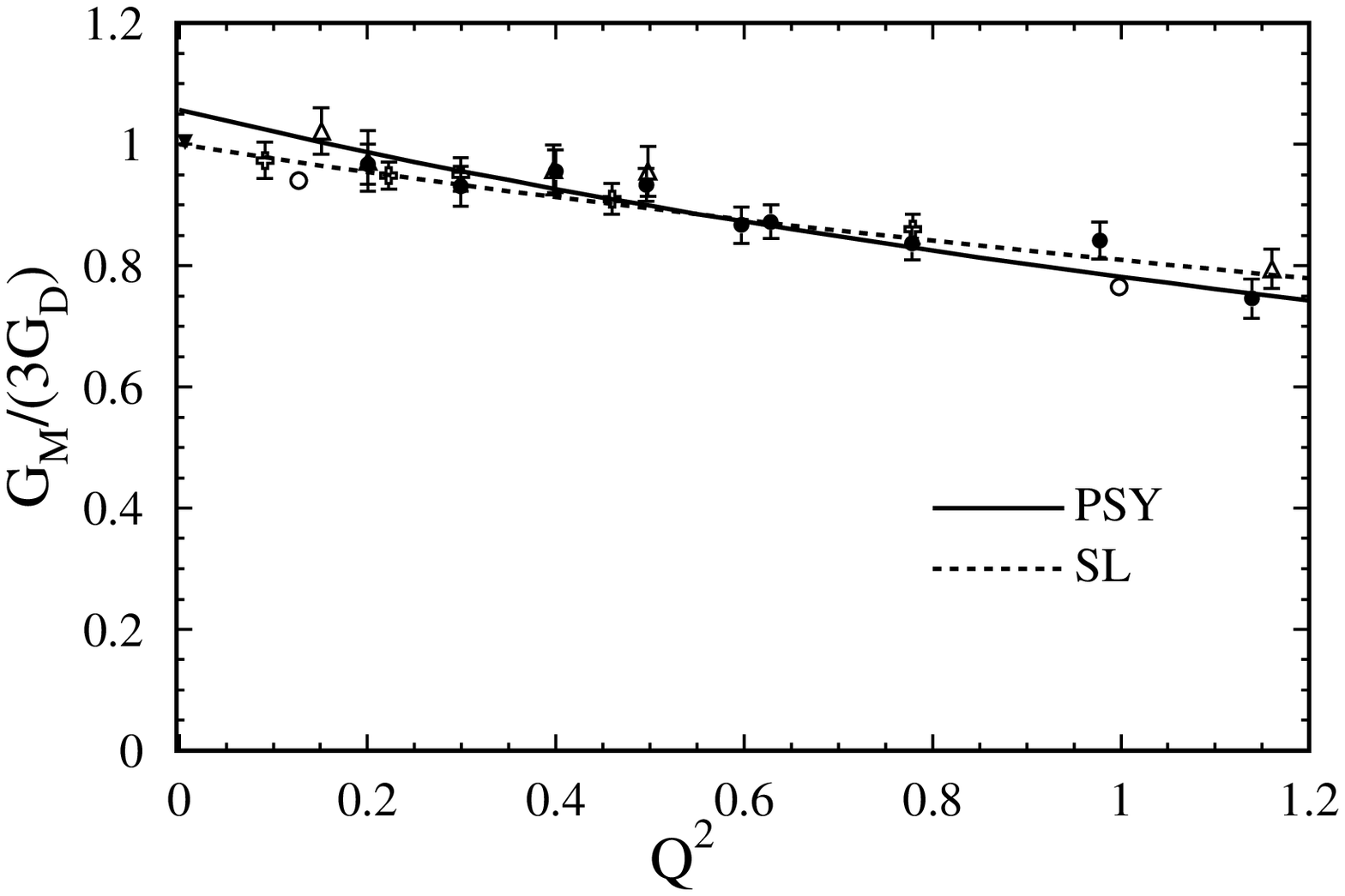}
\end{figure}
\begin{figure}[htb]
\centering
\vspace*{-2.5cm}
%\includegraphics[angle=0,width=8.5cm]{formf_fig1.eps}
%\vspace*{5.5cm}
\includegraphics[angle=0,width=8.5cm]{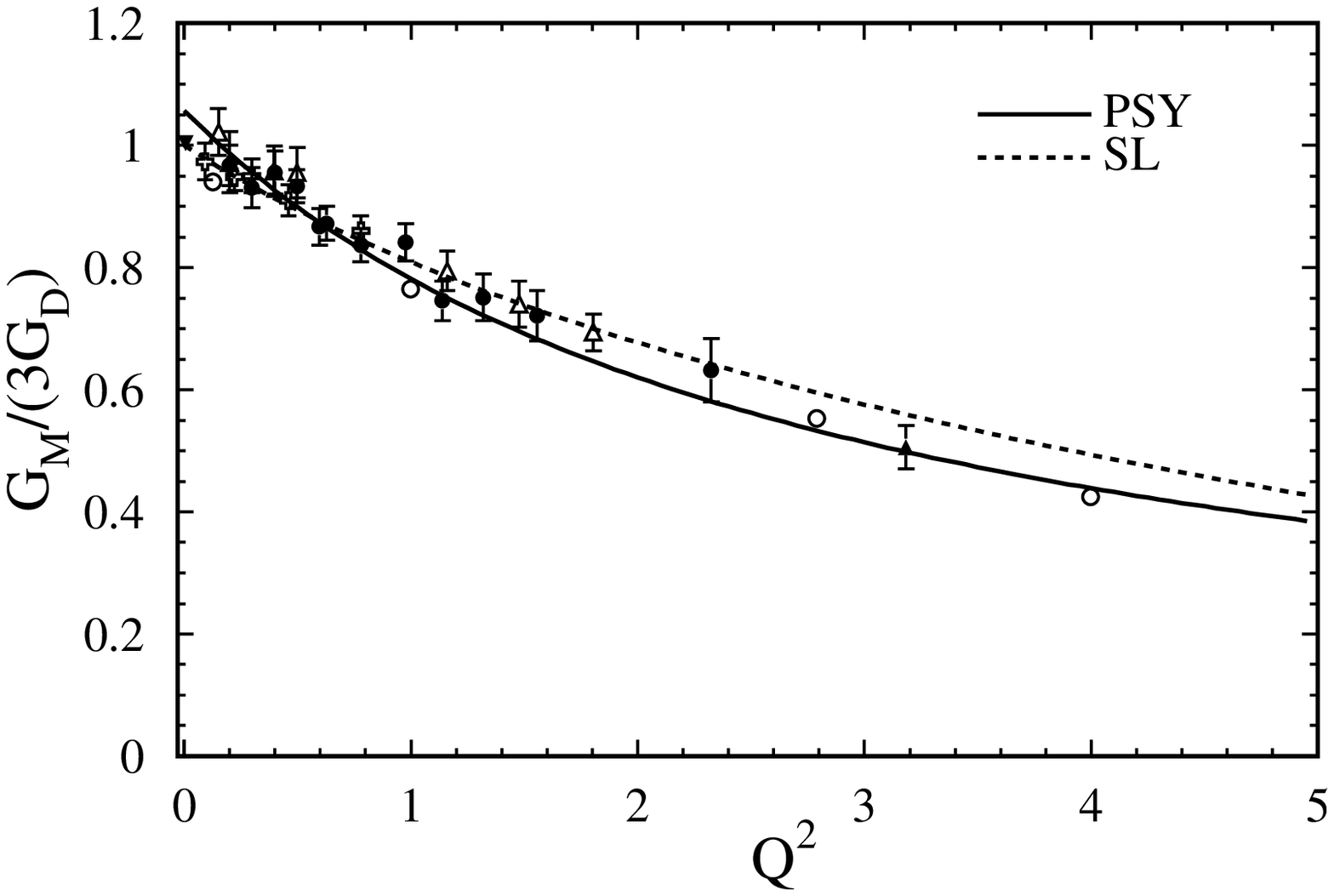}
\vspace*{-1.3cm}
\caption{\sf The magnetic $N-\Delta$ transition form factor 
$G_M^\star$ in dependence of $Q^2$ \cite{tit}. 
The solid and dotted lines denote 
our form factor eq.\ (\ref{eq:ff}) 
and the Sato-Lee(SL) form factor \cite{sl}, respectively.}
\label{tiator}
\end{figure}

The matrix element of the axial current has a similar
parameterization
\begin{eqnarray}
\langle \Delta^{++}|A_{\mu}|p\rangle 
&=& 
\Big[\bar{\psi}_{\mu} B_{\lambda}q^{\lambda}-
\bar{\psi}_{\lambda} q^{\lambda} B_{\mu}\nonumber\\
&+&
\bar{\psi}_{\mu}C_5^A +\bar{\psi}_{\lambda}
q^{\lambda}q_{\mu}C_6^A\Big]u(p)f(W)
\end{eqnarray}
with $B_{\lambda}=\frac{C_3^A}{M}\gamma_{\lambda}+
\frac{C_4^A}{M^2}p'_{\lambda}$.

The PCAC condition gives the relation
$C_5^A(q^2) = -\frac{C_6^A}{M^2}q^2$
which for small $q^2=0$ leads to the numerical value
$C_5^A(0) = 1.2$ \cite{ref5}.
The contribution of the form factor $C_6^A$ to the 
cross section is proportional to the lepton mass 
and will be ignored. 

The $Q^2$--dependence of the form factors varies among
the publications giving different cross sections and
different $Q^2$  distributions even when the same axial vector mass $M_A$
is used.  For this dependence we shall use a modified dipole form
\begin{equation}
C_5^A(Q^2) = \frac{1.2}{\left[1+\frac{Q^2}{M_A^2}\right]^2} 
\frac{1}{(1+\frac{Q^2}{3 M_A^2})}\ . \label{eq:ax}
\end{equation}
The proton has a charge distribution reflected in the form factor.
To build the resonance we must add a pion to the proton which creates 
a bound state with larger physical extent.
If the overlap of the wave functions has a larger mean-square-radius 
then the form factor will have a steeper $Q^2$ dependence 
as is indicated by the electromagnetic form factor for 
the excitation of $\Delta$ \cite{ref15}.
Since the effect is geometrical we expect a similar 
behavior for the vector and axial vector form factors. 
For this reason we replace another factor used in previous 
article \cite{ref5} by the modified dipole in eq.\ (\ref{eq:ff}) with 
$3 M_A^2 \sim 4 M_V^2 $.
For the other two form
factors $C_3^A(Q^2)$ and $C_4^A(Q^2)$ we shall use 
$C_3^A=0$ and $C_4^A(Q^2)=-\frac{1}{4}C_5^A$ \cite{ref5}.
It is evident that there is still arbitrariness
in the form factors with $C_3^A$ and $C_4^A$ being small.

\section{Numerical results} \label{neu}

\begin{figure}[htb]
\centering
\vspace*{-1.cm}
\includegraphics[angle=0,width=8.5cm]{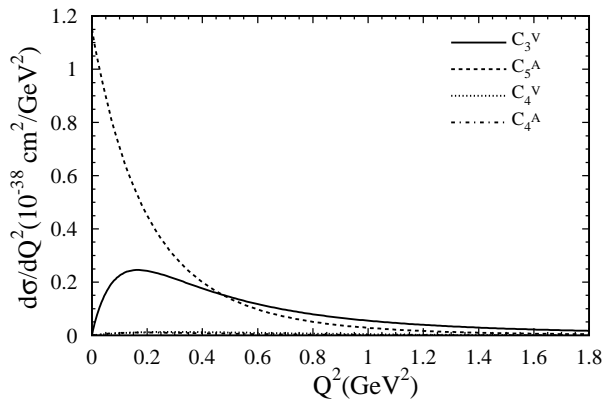}
\vspace*{-1.5cm}
\caption{\sf 
Individual contributions to the cross section separated by form factors. 
The solid and dotted curves are the contributions proportional to the 
vector form factors $C_3^V$ and $C_4^V$, respectively.
The dashed and dot-dashed curves denote the parts proportional to the 
axial vector form factors $C_5^A$ and $C_4^A$. }
\label{forf}
\end{figure}

We show in figure \ref{forf} the relative importance of 
the various form factors, where $C_3^V$ and $C_5^A$ dominate the cross section.
The cross section from the axial form factors 
has a peak at $Q^2= 0$, while the cross section from $C_3^V$ turns to zero.
The zero from the vector
form factor is understood, because in the 
configuration where the muon is parallel to the neutrino, 
the leptonic current is 
proportional to $q_\mu$ and takes the divergence
of the vector current, which vanishes by CVC.  
The contributions from
$C_4^V$ and $C_4^A$ are very small as shown in
the figure \ref{forf}.
For this reason the excitation of the $\Delta$ resonance, 
to the accuracy of present experiments, is well described 
by two form factors $C_3^V$ and $C_5^A$.

%Note that, since all form factors have been derived from photo- 
%and electroproduction experiments in which a $\Delta^+$ 
%or a $\Delta^0$ was produced, 
%all the form factors need to be multiplied by $\sqrt{3}$ 
%in order to obtain the correct cross section for the $\Delta^{++}$ 
%production due to 
%the fact that $<\Delta^{++}|V_\alpha|p> =\sqrt{3}<\Delta^+|V_{em}|p>$.

An estimate of the vector contribution is also possible 
using electroproduction data.
There are precise data for the electroproduction of 
the $\Delta$ and other resonances \cite{ref15}
including their decays to various pion-nucleon modes.
In the data of Galster et al.\ cross sections for the channels $(p+\pi^0)$
and $(n+\pi^+)$ are tabulated from which we conclude that both $I=3/2$ 
and $I=1/2$ amplitudes are present. 
For instance, for $W = 1.232\, {\rm GeV}$ the 
$I = 1/2$ background is $10\%$ of the cross section.

\begin{figure}[htb]
\centering
\vspace*{-1.cm}
\includegraphics[angle=0,width=8.5cm]{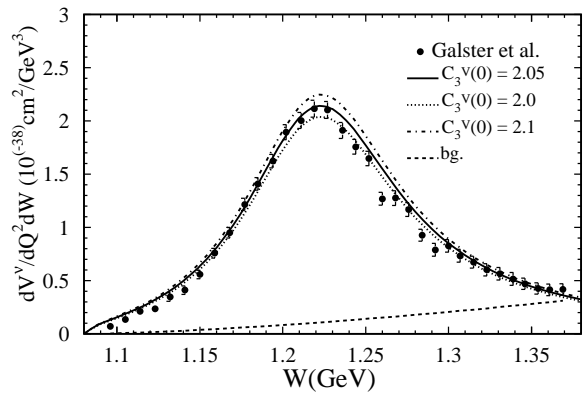}
\vspace*{-1.5cm}
\caption{\sf Cross section $dV^{\nu}/{dQ^2dW}$ for 
electroproduction in the $\Delta$ resonance in comparison with Galster et al.\
data \cite{ref15}. The solid, dotted and dot-dashed lines are obtained with
$C_3^V = 2.05$, $C_3^V= 2.0$ and $C_3^V= 2.1$, respectively. 
The dashed line denotes the backgroud contribution.}
\label{fig2}
\end{figure}

For our comparison in ref.\ \cite{ref10} we took the electroproduction 
data after subtraction of the background, as shown in fig. \ref{fig2}, 
and then used CVC to obtain the contribution of 
$V_\mu^+$ to neutrino induced reactions.
We used the formula
\begin{equation}
\frac{dV^{\nu}}{dQ^2dW} = \frac{G^2}{\pi}\,\,
\frac{3}{8}\,\, \frac{Q^4}{\pi\alpha^2}\,\,
\frac{d\sigma^{{\rm em},\, I = 1}}{dQ^2dW} \label{eq:e}
\end{equation}
to convert the observed \cite{ref15} cross sections 
for the sum of the reactions
$e + p \rightarrow  e + \left\{\begin{array}{c} p \,\pi^0\\ n \,\pi^+ 
\end{array} \right. $ to the vector contribution in the reaction 
$\nu +p \rightarrow \mu^- + p + \pi^+$ denoted in eq.\ (\ref{eq:e}) 
by $V^{\nu}$.
The factor $3/8$ originates from the Clebsch-Gordan 
coefficients relating the matrix elements of the two 
channels in the electromagnetic case to the matrix 
element of the weak charged current.
We used the data of Galster et al.\ \cite{ref15} at $Q^2 = 0.35 \,{\rm GeV}^2$
and subtracted the background as suggested by them.
Then we converted the points to the vector contribution 
for the neutrino reaction according to eq.\ (\ref{eq:e}).
In the same figure we show the 
neutrino cross section with $C_3^V(0) = 2.05$ (solid), 
$C_3^V(0) = 2.0$ (dotted), $C_3^V(0) = 2.1$ (dot-dashed) 
and contribution of background (dashed)   
and all other form factors equal to zero.
Before leaving this topic we mention that the analysis 
of the electroproduction data \cite{ref15} included a contribution from the 
$D_{13}(1520)$ resonance which was found to be small.

For the axial form factor we use the form given in eq.\ (\ref{eq:ax}). 
However, it is advisable to keep an open mind to notice 
whether a modification will become necessary.
With the method described here we have all parameters for the 
$\Delta$-resonance. We may still change the couplings by a 
few percent and vary $M_V$ and $M_A$. 

\begin{figure}[htb]
\centering
\vspace*{-1.cm}
\includegraphics[angle=0,width=8.5cm]{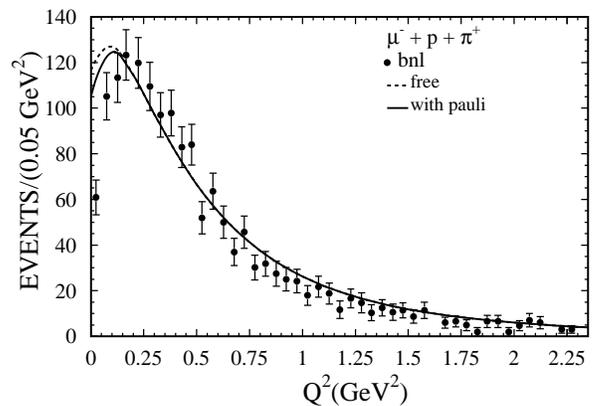}
\vspace*{-1.3cm}
\caption{\sf $Q^2$-spectrum of the process 
$\nu p \rightarrow \mu^- p \pi^+$ in comparison with BNL data 
\cite{ref19,furno}. The solid curve is with Pauli factor and dotted curve is free case.}
\label{fig3}
\end{figure}
There is still the $Q^2$ distribution \cite{ref18} 
to be accounted for. 
The data is from the Brookhaven experiment \cite{ref19,furno} 
where the experimental group  
presented a histogram averaged over the neutrino flux 
and with an unspecified normalization.
We used the form factors for the $\Delta$ resonance as introduced above
and included the correction from a Pauli factor 
in the simple Fermi gas model with Fermi momentum 
$p_F = 0.16\ {\rm GeV}$ \cite{ref10}.
The result is the solid curve 
shown in figure \ref{fig3} which is satisfactory. 
%for $Q^2\gtrsim 0.2\ \gevsq$.
For the relative normalization, we normalized the  area
under the theoretical curve for $Q^2 \geq 0.2 \ {\rm GeV}^2$ 
to the corresponding area under the data.
For the other parameters we choose
\begin{eqnarray}
&C_3^V(0) = 1.95\,
&, \quad C_5^A(0) = 1.2\, ,\nonumber\\
&M_V = 0.84 \,{\rm GeV}\,
&, \quad M_A = 1.05 \,{\rm GeV}\,.
\end{eqnarray}
In fact we made a $\chi^2$-fit and obtained these values 
with a $\chi^2$ per degree of freedom 
of 1.76 for the complete $Q^2$ region.
Furthermore, in order to reduce nuclear effects we performed a fit 
to all data with $Q^2>0.2\ {\rm GeV^2}$ and the fit result 
gave a $\chi^2/d.o.f = 1.04$.
In the theoretical curves we averaged over the neutrino 
flux for the BNL experiment \cite{baker}.  
The dotted curve is the calculation without Pauli factor 
and the solid one with Pauli suppression factor included 
which has a small effect.
It will be interesting to repeat this analysis when new data 
become available.
From fig.\ \ref{fig3} we can see that 
in the region of small $Q^2 \lesssim 0.2\ \gevsq$
the theoretical values are significantly above the experimental 
results which is not cured by a simple Pauli suppression factor
which generates small corrections.
Clearly, this discrepancy between data and theory is the origin
of the worse $\chi^2/d.o.f. = 1.76$ compared to the result of
$\chi^2/d.o.f. = 1.04$ 
where the low $Q^2$ region has been excluded.

\begin{figure}[htb]
\centering
\vspace*{-1.cm}
\includegraphics[angle=0,width=8.5cm]{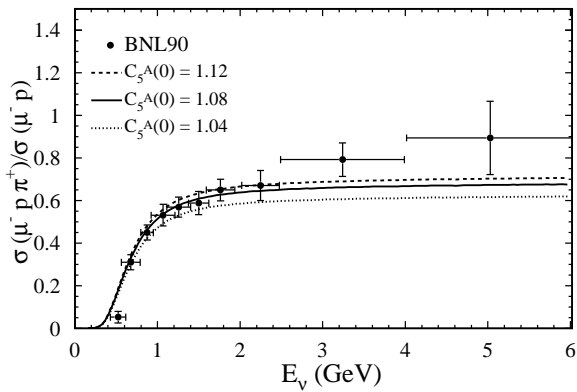}
\vspace*{-1.3cm}
\caption{\sf Ratio of total cross sections 
for RES and QE in comparison 
with BNL data \cite{ref19}.
The solid, dotted, and dashed lines have been obtained with 
$C_5^A = 1.08$, $C_5^A = 1.04$, and $C_5^A=1.12$, respectively.}
\label{fig4}
\end{figure}

\begin{figure}[htb]
\centering
\vspace*{-1.cm}
\includegraphics[angle=0,width=8.5cm]{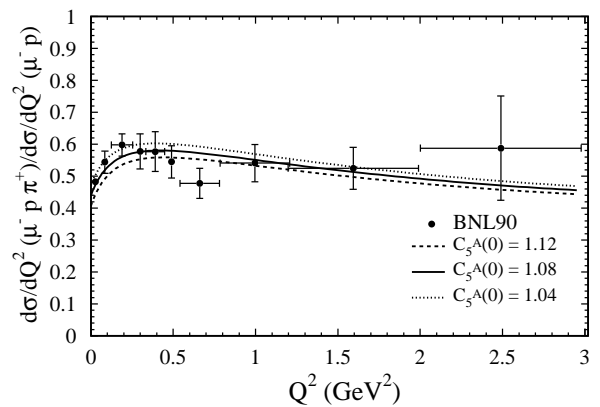}
\vspace*{-1.3cm}
\caption{\sf Ratio of $Q^2$ distributions for RES and QE
in comparison with BNL data \cite{ref19}. 
The lines are the same as in fig.\ \ref{fig4}. }
\label{fig5}
\end{figure}

For the resolution of this problem
it is reasonable to take the ratio 
of single pion production (RES) and quasi elastic scattering (QE) events,
$\frac{\sigma({\rm RES})}{\sigma({\rm QE})}$.  
In figs.\ \ref{fig4} and \ref{fig5} we show
this ratio for total cross sections and $Q^2$ distributions, respectively.
The curves have been calculated with
$M_A = 1.05\ {\rm GeV}$ and various values for $C_5^A(0)$ and 
have been compared with BNL data \cite{ref19}.
Since the ratio will reduce flux and experimental uncertainties
it is an especially good test for a theoretical model.
As can be seen from these figures we find very good agreement 
between our theoretical curves and the data, particularly also at low 
$Q^2 \lesssim 0.2\ {\rm GeV}^2$,  
because uncertainties in the low $Q^2$ region which are common to both,
resonance and quasi-elastic scattering, drop out in the ratios.

\begin{figure}[htb]
\centering
\vspace*{-1.cm}
\includegraphics[angle=0,width=8.5cm]{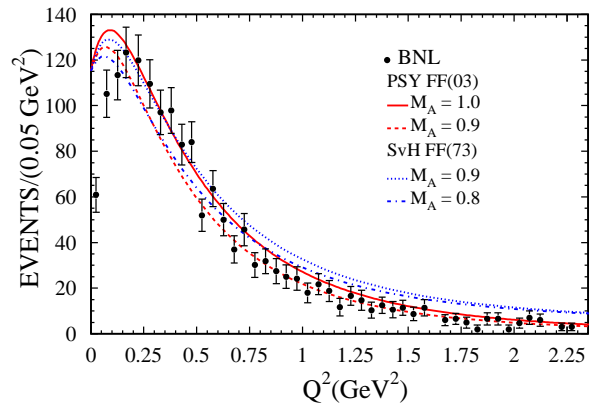}
\vspace*{-1.3cm}
\caption{\sf 
The same as in Fig.\ \ref{fig3} using for the theoretical
curves the PSY FFs \cite{ref10} (red, solid and dashed lines) 
and SvH FFs \cite{ref5} (blue, dotted and dot-dashed lines) 
for various values of the axial mass $M_A$.
}
\label{fig6}
\end{figure}
\begin{figure}[htb]
\centering
\vspace*{-1.cm}
\includegraphics[angle=0,width=8.5cm]{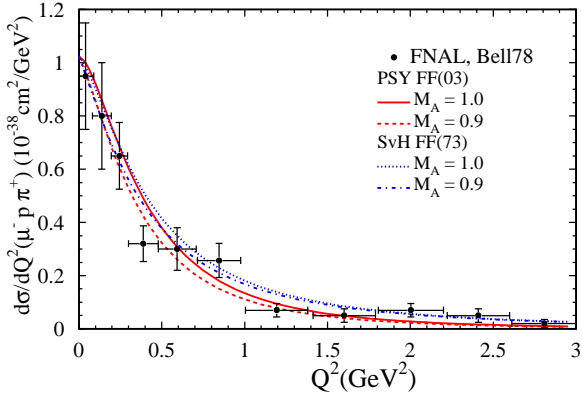}
\vspace*{-1.3cm}
\caption{\sf Same as in figure \ref{fig6} with FNAL data \cite{fnal}.}
\label{fig7}
\end{figure}
\begin{figure}[htb]
\centering
\vspace*{-1.cm}
\includegraphics[angle=0,width=8.5cm]{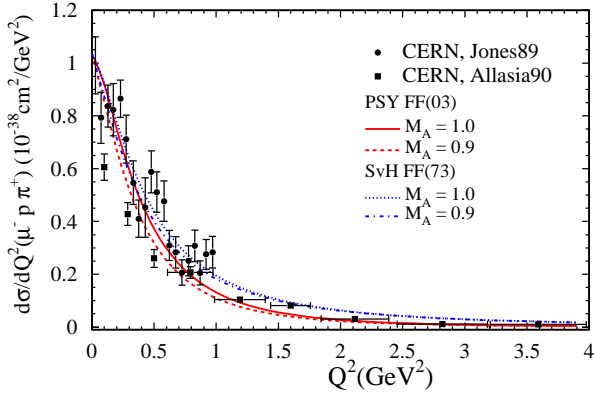}
\vspace*{-1.3cm}
\caption{\sf Same as in figure \ref{fig6} 
with CERN data \cite{cern}. }
\label{fig8}
\end{figure}

We now turn to a more detailed comparison of our 
form factors (PSY FFs) \cite{ref10} 
with form factors proposed by Schreiner and von Hippel
\cite{ref5} (SvH FFs) and additional experimental data for 
the $Q^2$ distribution of single pion production.
Figs.\ \ref{fig6}, \ref{fig7}, and \ref{fig8} show the flux averaged 
$Q^2$ spectrum for $\Delta$ resonance production in comparison 
with BNL, FNAL, and CERN data \cite{ref19,furno,fnal,cern}.
The theoretical curves have been plotted
using the PSY form factors (red, solid and dashed lines) 
and SvH form factors (blue, dotted and dot-dashed lines) 
for two $M_A$ values.
All these experimental results have been analyzed 
with a cut on the invariant mass, 
$W \leq 1.4\ {\rm GeV}$, such that the $\Delta$ resonance dominates.
One can see that the PSY FFs
lead to a better description of the data than the SvH FFs
for $Q^2 \gtrsim 1\ \gevsq$.
Also notice that the form factors become insensitive to $M_A$ 
for large $Q^2$ where $C_3^V$ (which does not depend on $M_A$) 
is dominant.

\section{$\Delta$ resonance electroproduction}\label{ele}

In this section we briefly discuss 
$\Delta$ resonance electroproduction.
We start with the fully differential cross section for neutrino reactions 
\cite{ref5} as follows
\begin{eqnarray}
\frac{{\rm d}^4 \sigma}{{\rm d}Q^2 {\rm d}W^2 {\rm d}\Omega_\pi^\star}
&=& \frac{1}{\sqrt{4\pi}} \frac{{\rm d}^4 \sigma}{{\rm d}Q^2 {\rm d}W^2} 
\Big( Y_0^0 
\nonumber\\
& &
- \frac{2}{\sqrt{5}}(\widetilde{\rho}_{(33)}-\frac{1}{2}) Y_2^0 
+ \frac{4}{\sqrt{10}}\widetilde{\rho}_{(31)} {\rm Re} Y_2^1 \nonumber\\
& &
-\frac{4}{\sqrt{10}}\widetilde{\rho}_{(3\ -1)} {\rm Re} Y_2^2 \Big)
\end{eqnarray}  
with the double differential cross section 
$\frac{{\rm d}^4 \sigma}{{\rm d}Q^2 {\rm d}W^2} 
= N \sum_{i=1}^3 K_i \widetilde{W}_i$ given in eq.\ (\ref{eq:douff}), 
$\widetilde{\rho}_{(33)},\ \widetilde{\rho}_{(31)},\ 
\widetilde{\rho}_{(3\ -1)}$ 
spin density matrix elements 
and $Y_l^m(\theta_\pi,\ \phi_\pi)$ spherical harmonic functions.
This cross section can be easily converted to the electroproduction
case by dropping axial vector parts and an appropriate change of
the normalization factor.
After some algebra we arrive at the following expression for 
the electroproduction 
which has a part proportional to $\cos2\phi_\pi^\star$ and
$\cos\phi_\pi^\star$ and a part independent of $\phi_\pi^\star$ 
\begin{eqnarray}
\frac{{\rm d}^4 \sigma}{{\rm d}Q^2 {\rm d}W^2 {\rm d}\Omega_\pi^\star}
&=& \frac{N}{\sqrt{4\pi}} \Big(\sum_{i=1}^3 K_i \big(\widetilde{W}_i - 
D_i \frac{(3\cos^2\theta_\pi^\star-1)}{2}\big)\nonumber\\
&-& 2\sqrt{3} \sin\theta_\pi^\star\cos\theta_\pi^\star\cos\phi_\pi^\star
(K_4 D_4 + K_5 D_5)\nonumber\\
&-& \sqrt{3} \sin^2\theta_\pi^\star\cos 2\phi_\pi^\star
(K_6 D_6) \Big)
\label{eq:electro}
\end{eqnarray}
with the normalization factor $N = \frac{\pi\alpha^2}{2 M^2Q^2}$,
$\theta_\pi^\star$, and $\phi_\pi^\star$ the pion polar and azimuthal angles,
respectively.
The kinematic factors
$K_i,\ i = 1,...,6$ and the structure functions
$\widetilde{W}_i, D_i$ can be found in \cite{ref5} and depend on
the vector form factors introduced in Sec.\ \ref{del}.

\begin{figure}[htb]
\centering
\vspace*{-0.3cm}
\includegraphics[angle=0,width=9cm]{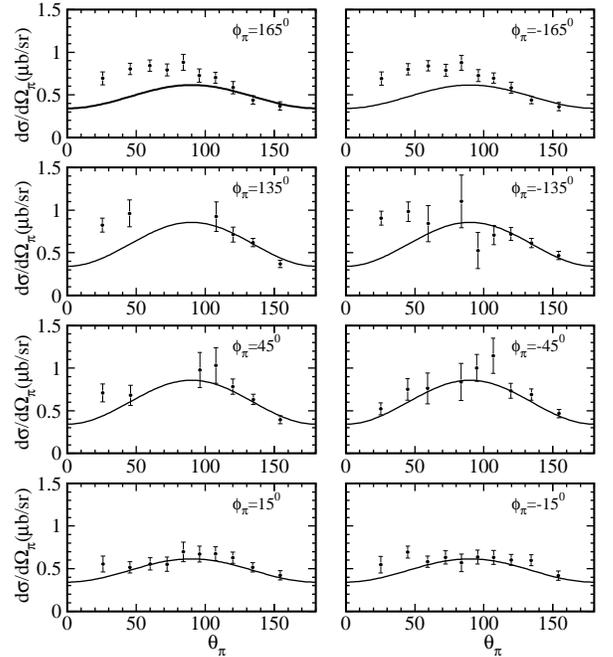}
\vspace*{-1.cm}
\caption{\sf Electroproduction data \cite{electro} for the reaction 
$e+p\rightarrow e+p+\pi^0$.}
\label{fig9}
\end{figure}
\begin{figure}[htb]
\centering
\vspace*{-0.5cm}
\includegraphics[angle=0,width=9cm]{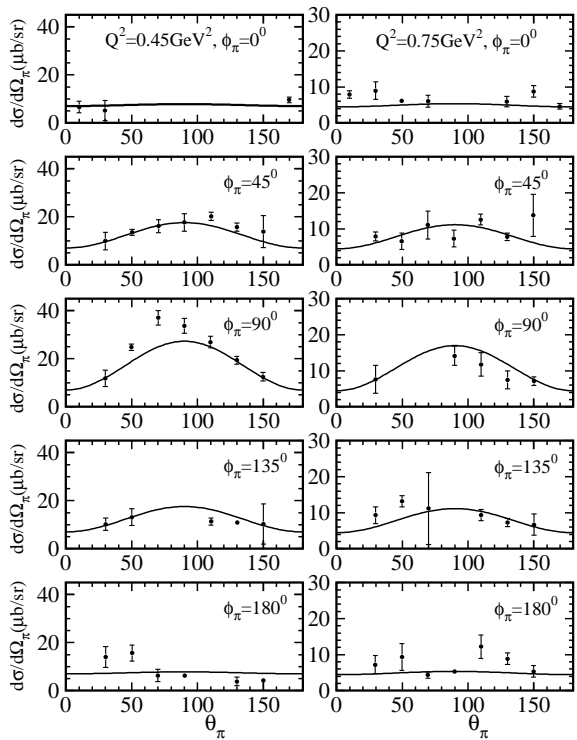}
\vspace*{-1.3cm}
\caption{\sf The same as in fig. \ref{fig9} for small 
$Q^2$ data \cite{siddle}.}
\label{fig10}
\end{figure}

For our numerical comparison with electroproduction data
we have calculated the fully differential
cross section in eq.\ \eqref{eq:electro} 
using the vector form factors given in eqs.\ \eqref{eq:c4v} and 
\eqref{eq:ff}.
In fig.\ \ref{fig9} we show electroproduction data \cite{electro} 
for the reaction $e + p\rightarrow e+p+\pi^0$ in dependence of the 
pion polar angle $\theta_\pi$ for various pion azimuthal 
angles $\phi_\pi$ for fixed 
$Q^2 = 2.8\ {\rm GeV}^2$, $W = 1.235\ {\rm GeV}$ and electron energy 
$E_e = 3.2\ {\rm GeV}$.
Similarly, in fig.\ \ref{fig10}
experimental results at small momentum transfers $Q^2= 0.45\ \gevsq$
and $Q^2= 0.75\ \gevsq$ are shown.
As can be seen the description of the data 
by our simple model
is already quite reasonable 
with exception of data points at $|\phi_\pi| \ge 135^\circ$
in the upper panels in fig.\ \ref{fig9}.
It should be noted that the theoretical curves have been obtained
without fine tuning of the parameter
$C_3^V(0)$, for which we took the value $C_3^V(0)=2.0$,
and without including the background from the $I=1/2$ channel
and a non-resonant background which is expected to be of the order
of $10 \%$.
A more detailed comparison will be presented elsewhere \cite{winp}.

\section{Conclusions}\label{summary}

%On the other hand, the comparison with existing neutrino- 
%and electroproduction data 
%is important in order to conform consistency checks of the theory and get 
%a deeper insight into the properties of resonances. 
%For this reason we will discuss how we relate the electroproduction 
%differential cross section to the one in neutrino production. 
%A detailed comparison with other theoretical models for electroproduction
%\cite{ref12} will be performed in the future \cite{winp}.
%We show that, at current experimental accuracy, the excitation 
%of the $\Delta$ resonance is described by two form factors,
%$C_3^V$ and $C_5^A$.
%Furthermore, relying on the connection of the weak vector current with
%the electromagnetic current due to CVC we compare 
%the vector part of our formalism
%with electroproduction data. 
Single pion neutrino production at low energies is important 
to measure properties of neutrino oscillations.
There exist already more complicated calculations, i.e., with 
more parameters 
and a more complicated functional form of the form factors. 
On the other hand, our approach is particularly simple and self-contained. 
Therefore it is helpful for the physical interpretation.

We have analyzed the differential cross section 
$\frac{{\rm d}\sigma}{{\rm d} Q^2}$ in terms of two form factors 
which are described by two free parameters $M_A$ and $M_V$ and 
compared our results with several experimental data. 
We obtained a good fit to BNL data \cite{furno} with a
$\chi^2/d.o.f = 1.04$ 
for $Q^2\geq 0.2 \ {\rm GeV}^2$. However, for $Q^2\leq 0.2\ {\rm GeV}^2$ 
the theory value is too high. A possible explanation are 
medium effects which are not properly taken into account.
The Pauli suppression factor has a small effect which does not
resolve the discrepancy. 
To shed light on this problem we took the ratio 
$\frac{\sigma {\rm (RES)}}{\sigma {\rm (QE)}}$ 
and found very good agreement with BNL data \cite{furno}.

From these comparisons we conclude that the neutrino production of the
$\Delta$ resonance can be described by two form factors $C_3^V$ and $C_5^A$.
$C_3^V$ is consistent with electroproduction data 
and $C_5^A$ can be determined only from neutrino data. 

We verified the consistency of our model by comparing
with further existing neutrino production data.
Moreover, we found that our form factors gave a better
description than the form factors originally proposed
by Schreiner and von Hippel \cite{ref5}.

Finally, we performed a first comparison with more differential 
electroproduction data and found reasonable agreement also here.
In this comparison we have neglected
the background
from the isospin $I =1/2$ channel and a non resonant background
which is of the order of $10 \%$ at $W=1232\ \gev$.

In the future 
we will include higher resonances and non-resonant 
background and perform a more detailed comparison 
with electroproduction data in order to study further the
connection between neutrino- and electroproduction.

\end{document}